\documentclass[aip]{revtex4-1}

\usepackage{amssymb,amsmath,amsfonts,amstext,mathtools}

\def\de{\delta}
\def\ep{\epsilon}
\def\ga{\gamma}
\def\vph{\varphi}
\def\th{\theta}
\def\bfmg{\mathbf{G}}
\def\bfmb{\mathbf{B}}
\def\bfme{\mathbf{E}}
\def\bfz{\mathbf{z}}
\def\bfp{\mathbf{p}}
\def\bfq{\mathbf{q}}
\def\bfmr{\mathbf{R}}
\def\sfb{\mathsf{b}}
\def\sfa{\mathsf{a}}
\def\sfc{\mathsf{c}}
\def\sfe{\mathsf{e}}
\def\hv{\mathsf v}
\def\p{\partial}
\def\na{\nabla}
\def\x{\times}
\DeclarePairedDelimiter\norm{\lVert}{\rVert}
\def\cdott{\mathord{\cdot}}
\def\bq{\begin{equation}}
\def\eq{\end{equation}}
\def\eqt#1{Eq.~(#1)}
\def\ncr{ \nonumber \\ }

\begin{document}

\title{On an intrinsic approach of the guiding-center anholonomy and gyro-gauge arbitrariness}

\author
{
L.~de~Guillebon}

\email{de-guillebon@cpt.univ-mrs.fr}

\author
{M.~Vittot
}

\affiliation{\makebox[0ex][l]{\small Centre de Physique Th\'eorique, Aix-Marseille Universit\'e, CNRS, UMR 7332,}\makebox[0ex][l]{\raisebox{-2.1ex}[0ex][0ex]{\small 13288 Marseille, France}}}

\affiliation{\makebox[0ex][l]{\small Universit\'e du Sud Toulon-Var, CNRS, UMR 7332, CPT, 83957 La Garde, France}\rule{0ex}{4.1ex}}

\begin{abstract}
In guiding center theory, the standard gyro-angle coordinate is associated with gyro-gauge dependence, the global existence problem for unit vectors perpendicular to the magnetic field, and the notion of anholonomy, which is the failure of the gyro-angle to return to its original value after being transported
around a loop in configuration space. We analyse these three intriguing topics through the lens of a recently proposed, global, gauge-independent gyro-angle. This coordinate is constrained, and therefore necessitates the use of a covariant derivative. It also highlights the intrinsic meaning and physical content of gyro-gauge freedom and anholonomy. There are, in fact, many possible covariant derivatives compatible with the intrinsic gyro-angle, and each possibility
corresponds to a different notion of gyro-angle transport. This observation sheds new light on Littlejohn's notion of gyro-angle transport and suggests a new derivation of the recently-discovered global existence condition for unit vectors perpendicular to the magnetic field. We also discuss the relationship between Cartesian position-momentum coordinates and the intrinsic gyro-angle.
\end{abstract}


\maketitle

	\section{Introduction}
	\label{intro}

 The dynamical behavior of a charged particle in a strong magnetic field exhibits a separation of time scales. These disparate time scales can be leveraged to isolate the slow guiding-center dynamics from the rapid Larmor motion, and to construct a constant of motion, the magnetic moment. This guiding-center reduction identifies a perturbative change of coordinates through an expansion in the Larmor radius to yield a slow ($4$-dimensional) description of charged-particle motion \cite{BogoZuba55, Krus62, NortAdia, NortRome78, Little81, Little83, CaryBriz09}. Guiding-center theory has wide applications in plasma physics, which involves space-time scales that are larger than the Larmor scales \cite{Bitt04, GoldRuth10}. In addition, it is the starting point for the gyrokinetic model of plasma dynamics, which is a key model in the study of plasma micro-turbulence \cite{BrizHahm07}. 
 
  The fast coordinate that measures the Larmor gyration of the particle around the magnetic-field lines is the so-called \textsl{gyro-angle}. It plays a leading role in guiding-center theory, but its standard definition raises several questions, both from a mathematical and from a physical point of view \cite{Little88, Sugi08, Krom09, Sugi09, BurbQin12}. For instance, when can the gyro-angle be globally defined? What is the physical and mathematical meaning of gyro-gauge dependence and gyro-angle anholonomy? These questions can imply troubles for guiding-center theory. For instance, at higher orders, the gauge-independence of the results can be difficult to prove, because it can rely on complicated vector identities \cite{BurbySquir13}. In addition, it was envisaged in the literature \cite{Sugi08} that the non-global existence might imply a breakdown of guiding-center theory at higher orders in the Larmor radius. Previous works clarified the point by investigating the condition for local descriptions to be consistent with a global description \cite{BurbQin12}, but an alternative and more radical way to solve the problem is to perform the derivation with a globally defined gyro-angle coordinate. 

 In recent works \cite{GuilMagMom, GuilGCmin}, we showed that both aspects of guiding-center theory (averaging reduction, and existence of the magnetic moment) can be addressed while using an alternative global, physical coordinate for the gyro-angle, which is the unit vector of the component of the momentum perpendicular to the magnetic field. No gauge fixing was needed. It was shown in Ref.~[\onlinecite{GuilMoyGa}] that this intrinsic coordinate can be used also in the standard procedure for the guiding-center reduction, which provides a Hamiltonian structure for the guiding-center dynamics and a maximal reduction for the guiding-center Lagrangian. All the results of the literature can thus be obtained using this physical but constrained coordinate. \\

 In the light of this approach, we now come back to the questions about the usual coordinate. This can clarify whether these questions are related to intrinsic properties of the physics and mathematics of the guiding-center system or to artificial, physically meaningless constructions. In the former case, it is interesting to study how these properties can be observed in the gauge-independent formulation, i.e. what the counterparts of the traditional questions are. Indeed, the guiding-center derivations in Refs.~[\onlinecite{GuilMagMom, GuilGCmin, GuilMoyGa}] showed that, in the gauge-independent approach, all the features of the standard approach seemed to be present, including a kind of generalized gauge vector. This makes it necessary to clarify whether the gauge-independent coordinate actually resolves the questions or only transfers them into other questions. 
 
 An additional advantage of the present study is to make clear the essential differences between the intrinsic gyro-angle and the traditional gyro-angle, which opens the way to variations of the intrinsic approach. For instance, this study could indicate how to eliminate the presence of a constrained coordinate while working with only intrinsically defined quantities, or how to wisely choose a scalar gyro-angle coordinate. 
 
 Thus, the purpose of this paper is twofold: first, the intrinsic counterparts of the guiding-center anholonomy and gauge arbitrariness are identified, and, second, we investigate whether it is possible to remove the intricacies of the intrinsic framework induced by its constrained coordinate system. \\

 The paper is organized as follows. In Sec.~II, we review the standard and the gauge-independent gyro-angles. This will emphasize that, when using the latter variable, the questions associated with the gyro-angle have disappeared from the coordinate system. Especially, the main difficulty, the non-global existence, will not have a counterpart in the intrinsic approach. 
 
 In Sec.~III, we turn our attention to the intrinsic counterpart of the gauge freedom. It is present in the intrinsic approach because the gauge-independent gyro-angle is treated as a constrained coordinate, which is not completely independent of the spatial position: when the spatial position is changed, the gyro-angle has to be changed in order to remain perpendicular to the magnetic field. This is similar to what happens to tangent vectors on a curved space, e.g. in general relativity. Thus, gradients are actually covariant derivatives. Their definition involves a free term, called the \textsl{connection} of the covariant derivative, which embodies the choice of a parallel transport for the gyro-angle. In the principal-circle-bundle picture \cite{Krus62, BurbQin12}, it corresponds to the connection $1$-form.
 
 In Sec.~IV, we investigate the anholonomy question. In the intrinsic framework, the anholonomy will result from the curvature of the coordinate space encoded in covariant derivatives, or more precisely in commutators between them. In the circle-bundle picture, it corresponds to the curvature $2$-form. 
 
 Secs.~II-IV are devoted to our first goal, namely the remnants of the three questions about the intrinsic coordinate system, while Secs.~V-VI are devoted to our second goal. These last two sections aim at simplifying the gauge-independent formalism by removing the presence of anholonomy and covariant derivatives along with their associated intricacies.
 
 In Sec.~V, we consider using the freedom embodied in the connection in order to remove the anholonomy by making covariant derivatives commute. This will provide an interesting approach to the existence condition for a scalar gyro-angle coordinate. 
 
 Lastly, in Sec.~VI, we consider the flaws associated with the presence of a constrained coordinate. We show how they can be eliminated by avoiding the splitting between the gyro-angle and the pitch-angle in the coordinate system.

	\section{A global gauge-independent coordinate for the gyro-angle}

 The physical system under consideration is a charged particle with position $\bfq$, momentum $\bfp$, mass $m$, and charge $e$, under the influence of an electromagnetic magnetic field $(\bfme, \bfmb)$. The motion is given by the Lorentz force
\begin{align}
	\dot \bfq &= \tfrac {\bfp}{m}
	\,,\ncr
	\dot \bfp &= \tfrac {\bfp}{m}\x e\bfmb+e\bfme
	\,.\notag
\end{align}

 When the magnetic field is strong, the motion implies a separation of time scales. This is best seen by choosing convenient coordinates for the momentum space, for instance \cite{GuilGCmin}
\begin{align}
	p &:= \norm \bfp
	\,,\ncr
	\vph &:= \arccos \left(\tfrac{\bfp\cdott\sfb}{\norm{\bfp}}\right)
	\,,
\label{GyroangleIntrDef}
\\
	\sfc &:= \tfrac{\bfp_\perp}{\norm{\bfp_\perp}}
	\,,\notag
\end{align}
where $\sfb:= \tfrac{\bfmb}{\norm{\bfmb}}$ is the unit vector of the magnetic field, and $\bfp_\perp:=\bfp-(\bfp\cdott\sfb)\sfb$ is the so-called \textsl{perpendicular momentum}, i.e. the orthogonal projection of the momentum onto the plane perpendicular to the magnetic field. The coordinate $p$ is the norm of the momentum. The coordinate $\vph$ is the so-called \textsl{pitch-angle}, i.e. the angle between the momentum and the magnetic field. The last variable $\sfc$ is the unit vector of the perpendicular momentum.

 Then, the equations of motion write
\begin{align}
	\dot \bfq &= \tfrac {\bfp}{m}
\,,\ncr
	\dot p & = \tfrac{e\bfme\cdott\bfp}{p}
\,,\ncr
	\dot \vph & = -\tfrac{\bfp}{m}\cdott\na\sfb\cdott\sfc
	+\tfrac{e\bfme}{p\sin\vph}
	\cdott\left(\cos \vph ~\tfrac{\bfp}{p} - \sfb\right)
	\,,
\label{Dynamics}
\\
	\dot \sfc & = - \tfrac{e B}{m}\sfa 
	- \tfrac{\bfp}{m}\cdott\na\sfb\cdott(\sfc\sfb 
							+ \sfa\sfa \cot \vph )
	+ \tfrac{e \bfme\cdott \sfa}{p ~ \sin\vph}\sfa
	\,,\notag
\end{align}
where $\bfp$ is now a shorthand for $p(\sfb \cos \vph + \sfc \sin \vph)$, $B$ is the norm of the magnetic field and, following Littlejohn's notations \cite{Little81, Little83}, the vector $\sfa:=\sfb\x\sfc$ is the unit vector of the Larmor radius, so that $(\sfa,\sfb,\sfc)$ is a right-handed orthonormal frame (rotating with the momentum). 

 In the case of a strong magnetic field, the only fast term is the Larmor frequency $\omega_L:=\tfrac{e B}{m}$, which corresponds to the gyration of the particle momentum around the magnetic field. It concerns only one coordinate, namely $\sfc$, the direction of the perpendicular momentum $\bfp_\perp$ in the $2$-dimensional plane perpendicular to the magnetic field. This coordinate corresponds to the gyro-angle. \\

 To get a scalar angle instead of the vector $\sfc$, one chooses at each point $\bfq$ in space a direction $\sfe_1(\bfq) \in \bfmb^\perp(\bfq)$ in the plane perpendicular to the magnetic field, which will be considered as the reference axis. Then, the angle $\theta$ is defined as the oriented angle between the chosen reference axis $\sfe_1(\bfq)$ and the vector $\sfc$ through the following relation~:
\bq
	\sfc = - \sfe_1 \sin \theta - \sfe_2 \cos \theta 
	\,,
\label{GyroangleThetaDef}
\eq
with $\sfe_2:=\sfb\x\sfe_1$ the unit vector such that $(\sfb,\sfe_1,\sfe_2)$ is a (fixed) right-handed orthonormal frame \cite{Little81}. The angle $\th$ is the usual coordinate for the gyro-angle \cite{BogoZuba55, NortAdia, Little81, CaryBriz09}. Its equation of motion is
\bq
	\dot \theta = 
	\tfrac{e B}{m} 
	+ \cot \vph \tfrac{\bfp}{m}\cdott\na\sfb\cdott\sfa 
	+ \tfrac{\bfp}{m}\cdott\na\sfe_1\cdott \sfe_2
	- \tfrac{e \bfme\cdott \sfa}{p ~ \sin\vph}
	\,.
\label{DynamicsTheta}
\eq

From the initial dynamics $(\dot\bfq, \dot p,\dot \varphi, \dot \theta)$, guiding-center reductions perform a change of coordinates $(\bfq, p, \varphi, \theta)\longrightarrow (\bar\bfq, \bar p, \bar\varphi, \bar\theta)$ in order to obtain a reduced dynamics with suitable properties, mainly a constant of motion $\dot{\bar p}=0$ and a slow reduced motion $(\dot{\bar\bfq}, \dot{\bar\varphi})$ that is both independent of the fast coordinate $\bar\theta$ and Hamiltonian \cite{CaryBriz09, GuilMoyGa}. The reduced position $\bar\bfq$ is the \textsl{guiding-center}. The constant of motion $\bar p$ is the \textsl{magnetic moment}. It is close to the well-known adiabatic invariant $\mu:=\tfrac{\bfp_\perp^2}{2mB}$, and is usually written $\bar\mu$ instead of $\bar p$. \\

 In the definition of $\th$, the necessary introduction of $\sfe_1(\bfq)$ implies important and awkward features in the theory. First, the choice of $\sfe_1(\bfq)$ is arbitrary, which induces a local gauge in the theory. The coordinate system is gauge dependent since the value of $\th$ depends of the chosen $\sfe_1(\bfq)$. For a general reduction procedure, the guiding-center dynamics can end up being gauge dependent. For instance, the maximal reduction by Lie-transforming the phase-space Lagrangian is gauge dependent \cite{GuilMoyGa}.
 
 Guiding-center reductions have to use prescriptions in order to avoid such unphysical results. For instance, in the reduced Lagrangian \cite{Little83, CaryBriz09}, the 1-form d$\th$ must appear only through the quantity $d\th-(d\bfq\cdott\na\sfe_1)\cdott\sfe_2$. These gauge-dependence questions emphasize that the gyro-angle is artificial, it is not given by the physics and its meaning is restricted.
 
 Second and more substantial, a continuous choice of $\sfe_1(\bfq)$ does not exist globally in a general magnetic geometry \cite{Sugi09, BurbQin12}. The reason is that the possible values for $\sfe_1(\bfq)$  define a principal circle bundle over the configuration space \cite{Krus62, BurbQin12}. A specific choice $\sfe_1(\bfq)$ is a global section of the bundle. Such a global section induces a trivialization of the bundle, whose coordinate system is precisely $(\bfq,\th)$. But a trivialization does not exist globally for a general circle bundle. In the case of the guiding-center, this global non-existence can be proven by using the theory of principal bundles and characteristic classes \cite{BurbQin12}. Thus, the gyro-angle does not exist in the whole physical system in general. The coordinate system $(\bfq, p, \varphi, \th)$ does not capture the mathematical description of the system, only a strongly simplified description that is valid only in the trivial case. 
 
 It was mentioned in Ref.~[\onlinecite{BurbQin12}] that the local descriptions are consistent with a global description provided the change of local descriptions satisfy some relations. For instance, the $1$-form $d\theta$ must appear only through the combination $d\th-(d\bfq\cdott\na\sfe_1)\cdott\sfe_2$. So, working with $\th$ is not meaningless. For instance, it can provide a slow guiding-center dynamics $(\dot{\bar\bfq}, \dot{\bar\vph})$ that is globally defined. Nevertheless, this does not provide the local coordinate $\th$ with a global meaning, neither does it explain what the global description of the gyro-angle is. 
 
 Last, even the local description is not completely regular, because it involves a non-holonomic phase in the gyro-angle. When a loop $\ga$ is performed in position space (while keeping the momentum coordinates $(p, \varphi, \theta)$ constant), at the end of the process all the coordinates and all the physical quantities have recovered their value, but the variation of the gyro-angle involves non-zero partial contributions \cite{Little81, Little88, Sugi08}. The best known of these contributions is the so-called {\textsl geometric phase} $\Delta\theta_g$. It is related to the third term in the right-hand side of \eqt{\ref{DynamicsTheta}}: 
\begin{equation}
	\Delta\theta_g
	:=
	\oint_\ga (d\bfq\cdott\na\sfe_1)\cdott\sfe_2
	\,.
\label{GeomPhaseIni}
\end{equation}
It embodies the well-known anholonomy associated with to the gyro-angle coordinate. 
 
 A similarity with Berry's phase and more generally with Hannay's phase was often pointed out \cite{Little81, Little88, LiuQin11}, but there are significant differences, as mentioned in Ref.~[\onlinecite{Little88}]. Especially, these phases are related to adiabaticity with a single path in parameter space followed by the system, which makes them physically determinable. On the contrary, guiding-center anholonomy is related to path dependence in configuration space, with all paths coexisting simultaneously. This precludes any definite value for this phase. It raises questions whether this phase is physically meaningful or if it affects only non-physical quantities concerning the extrinsic coordinate system. \\

 All these questions arise because $\th$ is only an artificial quantity. This fact motivated to keep the primitive gauge-independent coordinate $\sfc$ instead of introducing $\th$. Even if this quantity does not have scalar values, it embodies an angle since it is a unit vector in a plane and hence belongs to a circle $\mathbb S^1$. 
 
 The vector $\sfc$ represents the physical quantity corresponding to the gyro-angle; $\th$ never appears alone in the theory (e.g. in guiding-center transformations), except in its own definition and subsequent relations. What appears everywhere is the vector $\sfc$ (see e.g. Refs.~[\onlinecite{NortAdia, Little81, CaryBriz09}]). Even for the correction to the gyro-angle $(\bar\th-\th)$ in guiding-center reductions, $\th$ never appears alone but only as the argument of the vector $\sfc$. An example can be found in the equations of motion (\ref{Dynamics}) and (\ref{DynamicsTheta}), which also illustrate the gauge dependence or independence.

 In  addition, the coordinate $\sfc$ is globally defined, since the perpendicular momentum is well defined everywhere in the guiding-center system. It is useful to remember that the points where the momentum is parallel to the magnetic field (i.e. the points where $\varphi=0$) are always implicitly excluded from guiding-center theory, even in the local description using the scalar coordinate $\th$. Indeed, at those points, the angle $\th$ cannot be defined. In addition, guiding-center expansions involves some $\sin\varphi$ in denominators \cite{SingularityGC}, which implies to exclude the points where $\varphi=0$.

 The variations of $\sfc$ do not include the gauge contribution $(d\bfq\cdott\na\sfe_1)\cdott\sfe_2$, with its anholonomic geometric phase. In addition, no extrinsic effect can be observed in the coordinate system. So, one can expect the questions about the anholonomic geometrical phase to disappear from the theory. The underlying idea is basically true, as will be confirmed by the next sections, but some subtleties should be taken into account. As is emphasized in Eq.~(\ref{GeomPhaseIni}), anholonomy does not affect the coordinate system itself, but rather some contributions to the gyro-angle after one loop in phase space (or in configuration space with a given parallel transport for the gyro-angle). The analysis in the next sections will show that the coordinate system is not enough to define the corresponding contributions, but that they can indeed be defined with a zero geometric phase. All the same, anholonomy will not disappear from the system. It will be induced by the structure of the coordinate system, and especially by the magnetic geometry. Accordingly, the coordinate $\sfc$ does not remove anholonomy but it does remove the anholonomy question: with the coordinate $\th$ the puzzling point was not anholonomy in itself, but rather the fact that in principle anholonomy is absent when the coordinate system is trivial.
 
 Last, the coordinate $\sfc$ agrees with the mathematical description of the system. For any magnetic geometry, it induces a circle bundle \cite{Little88,BurbQin12,Krus62}. Indeed, the circle for $\sfc$ is position-dependent since $\sfc$ is perpendicular to the magnetic field. So, $\sfc$ is not just in $\mathbb S^1$ but in $\mathbb S^1(\bfq)$. A few consequences will be studied in the next sections. In the traditional coordinate system, this picture is absent because $\th$ is independent of the position. The circle bundle rather concerns the vector $\sfe_1$, but the corresponding bundle is different from the intrinsic bundle for the gyro-angle. It is not defined by the whole phase space but confined to the four-dimensional space $(\bfq, \sfe_1)$. In addition, the global section $\sfe_1$ assumes the topology of the bundle trivial. 

 Accordingly, the use of the gyro-angle $\sfc$ removes from the coordinate system all of the questions involved in the standard gyro-angle $\th$. It indeed provides the intrinsic description of the physics and mathematics of the system.

\section{Intrinsic counterpart of the gauge arbitrariness}

 The previous section showed that in the intrinsic framework, the gauge arbitrariness disappears from the coordinate system. In this section, we show that some arbitrariness remains in the theory and that it corresponds to the intrinsic counterpart of the gyro-gauge arbitrariness
 
 Indeed, when the coordinate $\sfc$ is used, the spatial dependence of $\mathbb S^1$ causes the coordinate space to be constrained, i.e. the coordinates are not completely independent of each other. When the position $\bfq$ is changed, the coordinate $\sfc$ cannot be kept unchanged, otherwise it may get out of $\sfb^\perp$:
\bq
	\na\sfc\neq0
	\,.
\eq
Differentiating \eqt{\ref{GyroangleIntrDef}} with respect to $\bfq$, one finds \cite{GuilGCmin}
\bq
	\p_{\bfq|\bfp} \sfc 
	= - \na\sfb\cdott(\sfc\sfb+\sfa\sfa\cot\vph)
	\,.
\label{ConnectNatur}
\eq
The right-hand side is well defined everywhere, since the points where $\cot\vph=\pm\infty$, i.e. where $\bfp$ is parallel to the magnetic field $\bfmb$, are excluded from the theory. \\

 \eqt{\ref{ConnectNatur}} must not be given a completely intrinsic meaning, because the two terms in its right-hand side play a very different role through coordinate change: the first term is always unchanged, whereas the second one is generally changed. For instance, if one uses the scalar angle $\th$ as a local coordinate for $\sfc$, \eqt{\ref{ConnectNatur}} becomes
\bq
	\na\sfc 
	= - \na\sfb\cdott\sfc~\sfb +\bfmr ~\sfa
	\,,
\label{ConnectGauge}
\eq
where 
\bq
	\bfmr:=\na\sfe_1\cdott\sfe_2
\eq
is the so-called \textsl{gauge vector}. $\bfmr$ is a function of the position $\bfq$, and it is not unique: it depends of the choice of gauge $\sfe_1(\bfq)$. 

 The reason for this difference of role is that the definition space $\mathbb S^1(\bfq)$ for the coordinate $\sfc$ exactly imposes the first term in \eqt{\ref{ConnectNatur}} but gives no constraints on the second term. Indeed, the gyro-angle $\sfc$ is a free $1$-dimensional coordinate, but it is at the same time a vector immersed in $\mathbb R^3$. The two remaining dimensions are fixed by the condition for $\sfc$ to have unit norm $\sfc\cdott\sfc=1$ and to be transverse to the magnetic field $\sfc\cdott\sfb=0$. This implies 
\bq
	\na\sfc\cdott\sfc=0
	\text{~~~~and~~~~}
	\na\sfc\cdott\sfb=-\na\sfb\cdott\sfc
	\,.
\eq

 Thus, in $\na\sfc$, only the component parallel to $\sfa$ is not imposed by intrinsic properties linked with $\mathbb S^1(\bfq)$. Practically, it is induced by the specific definition chosen for the gyro-angle coordinate. From an intrinsic point of view, it is completely free:
\bq
	\na\sfc 
	= - \na\sfb\cdott\sfc~\sfb+\bfmr_g~\sfa
	\,,
\label{ConnectGene}
\eq
where 
\bq
	\bfmr_g:=\na\sfc\cdott\sfa
\eq
is a free phase-space function.\\

 The geometric picture of this freedom is the following. In the gradient $\na\sfc$, i.e. in the effects of an infinitesimal spatial displacement on $\sfc$, one of the terms, $- \na\sfb\cdott\sfc~\sfb$, is mandatory: it is necessary and sufficient for $\sfc$ to remain inside its definition domain in the process of spatial transportation. This is easily seen on a diagram. The other term $\na\sfb\cdott\sfa~\sfa$ is only optional: it corresponds to a rotation of $\sfc$ around $\sfb$ (hence a gyration around the circle) accompanying the spatial displacement. However, this term and its associated rotation could be removed or given a different value. They are not imposed by intrinsic properties and have to be arbitrarily chosen, in a similar way as when the target set of a projection is determined, but not the kernel. They encode the way points in a circle $\mathbb S^1(\bfq_1)$ at the position $\bfq_1$ are "projected" (more precisely connected) to points in the circle $\mathbb S^1(\bfq_1+\delta \bfq_1)$ at a neighbouring position. 
 
 This is very similar to what occurs to tangent vectors on a curved space, for instance on the surface of the Earth or in general relativity. When the spatial position is moved, the tangent vectors from the initial point have to be moved into the tangent space of the final point. Otherwise they are not tangent vectors any more. So, the gradient is actually a covariant derivative, that is an infinitesimal operator that not only changes the position, but also all other quantities in such a way that they remain inside their definition domain throughout the displacement. This phenomenon has to appear as soon as the coordinate space is constrained, i.e. when the definition domain of some coordinate (or parameter) depends on other coordinates. 
 
 As a result, when using the coordinate $\sfc$, the generator of spatial displacements is not a standard gradient operator, such as the gradient with the gauge-dependent gyro-angle coordinate or the one with the Cartesian position-momentum coordinate. In order to emphasize the difference, the covariant-derivative gradient will be denoted with an over-bar $\bar \na$. When acting on functions depending only on $\bfq$, the operators $\na$ and $\bar\na$ are the same, and the over-bar will not be used. 
 
 In the definition of the covariant derivative, there is always some freedom, since the definition domain at the initial point can be connected to the definition domain at the final point in an arbitrary way. The free term in the covariant derivative corresponds to its \textsl{connection}. It determines the way objects are parallel transported. In the example of tangent vectors on a curved surface, the so-called affine connection corresponds to the well-known Christoffel symbols. For the gyro-angle $\sfc$, the connection freedom is embodied in $\bfmr_g$.

 More generally, the connection freedom might not affect only gradients $\na$. Other derivative operators could also become covariant derivatives impacting the gyro-angle, i.e. 
\begin{equation}
\partial_p\sfc=f_1~\sfa
\text{~~ and ~~}
\partial_\varphi\sfc=f_2~\sfa
\,,
\label{ConnectGeneralized}
\end{equation}
with $f_1$ and $f_2$ arbitrary functions of the phase space. However, this additional refinement will not be used here, because it does not seem to be useful nor natural: the definition space for the gyro-angle $\sfc$ depends only on the magnetic field, and hence of the spatial coordinates. 

 All these features agree with the circle-bundle picture and emphasize its relevance, which was hidden and even suppressed from the coordinate system when using the gauge-dependent gyro-angle. Indeed, instead of viewing the phase space as a constrained coordinate system $(\bfq, p,\varphi, \sfc)$ as we did above, one can equivalently consider it as a principal $\mathbb S^1$-bundle with a base space given by $(\bfq, p, \varphi)$ and a fiber $\mathbb S^1$. This setting offers rigorous mathematical structures. Then, the covariant derivative is defined in terms of the parallel transport associated to a chosen connection $1$-form, i.e. a $1$-form on the fiber bundle that evaluates to $1$ on the canonical infinitesimal generator of gyro-rotation (see e.g. Ref.~[\onlinecite{HamilRedStag}] for an introduction to the subject). In the present paper, we prefer to avoid focusing on this more technical viewpoint. Its terminology and concepts could seem unfamiliar to some readers, and they are not indispensable to investigate the guiding-center system, which can be considered in the more basic picture of a constrained coordinate system. 

 In either formalism, the conclusion is that, although the gyro-gauge with its arbitrariness is absent from the intrinsic approach, some arbitrariness is present, not in the coordinate system itself but in the choice of a connection for the covariant derivative. \\

 The comparison with the gauge-dependent approach emphasizes the relationship between the two arbitrarinesses. \eqt{\ref{ConnectGauge}} shows that the connection is then embodied in the gauge vector $\bfmr$. This quantity is called gauge vector \cite{Little81, Little88} because it exactly embodies the information on the local gauge. Changing the gauge $\sfe_1(\bfq)\longrightarrow \sfe_1'(\bfq)$ just means adding to the value of $\th$ a scalar function $\psi(\bfq$), whose value is the angle between $\sfe_1$ and $\sfe_1'$ at $\bfq$. Under this change, the gauge vector is changed by 
\bq
	\bfmr'=\bfmr+\na\psi
	\,.
\label{GaugeVectGaugeDepend}
\eq
The term $\na\psi$ exactly embodies the purely local gauge. The remaining part of the gauge corresponds to a global gauge, i.e. the value of $\psi$ at a reference point $\bfq_1$. Changing it corresponds to a global rotation by the same angle $\psi(\bfq_1)$ for all points $\bfq$. It is not specifically a (local) gauge, but rather a common change of angular coordinate. 

 As a side comment, which will be useful for the following, \eqt{\ref{GaugeVectGaugeDepend}} shows that, even if the gauge vector is gauge dependent, its curl is not. The curl \cite{Little81}
\begin{align}
	\na\x\bfmr 
	=&\mathbf N 
	\,,
\label{ConditionIntegrGaugeVect}
\\
	\text{~~with~~} 
	\mathbf N: 
	=& 
	\tfrac{\sfb}{2}
	\big(
	\text{Tr}(\na\sfb\cdott\na\sfb)
	-
	(\na\cdott\sfb)^2
	\big)
\ncr
	& +(\na\cdott\sfb)\sfb\cdott\na\sfb
	-
	\sfb\cdott\na\sfb\cdott\na\sfb
\label{VectCurlGaugeVect}
\end{align}
is related to intrinsic properties of the system. Accordingly, the gauge vector cannot be a free function of the position space, but it must satisfy the integrability condition (\ref{ConditionIntegrGaugeVect}). Especially, the choice $\bfmr=0$ is not available in general, even locally. \\

 With the above interpretation for $\bfmr$, \eqt{\ref{ConnectGauge}} means that the local gauge freedom is exactly the counterpart of the connection freedom $\bfmr_g$. The geometric reason is that $\th$ is the angle between $\sfe_1$ and $\sfc$. Through spatial transportation $\de\bfq$ with $\th$ unchanged, the gauge vector $\bfmr$ means that $\sfe_1$ is rotated around $\sfb$ by an angle $\de\bfq\cdott\bfmr$. It implies that $\sfc$ undergoes the same rotation. Thus, the connection for $\sfe_1$ is at the same time a connection for $\sfc$.
 
 This discussion shows that the gyro-gauge freedom originates from an inherent property of the guiding-center system. It also clarifies the precise meaning and physical content of this intrinsic arbitrariness in two ways. First, it indicates that the arbitrariness is not associated with a choice of coordinate system, but rather a choice of covariant derivative. Second, it implies not only a restricted class of functions of the configuration space but a free function of the phase space. 
 
 The first point clarifies how in the intrinsic formulation the arbitrariness is present but does not affect the coordinates at all, whereas in the standard formulation it affects also the coordinate $\th$. A first illustration is provided by the guiding-center transformation. For the gyro-angle $\sfc$, the transformation is connection-independent \cite{GuilGCmin, GuilMoyGa}. For the coordinate $\th$, the transformation $\bar\th=\cdott\cdott e^{\bfmg_2}e^{\bfmg_1}\th$ is gauge dependent, but in such a way as to make the induced transformation $\bar\sfc=\cdott\cdott e^{\bfmg_2}e^{\bfmg_1}\sfc$ for $\sfc$ gauge independent, with $\bfmg_n$ the vector field generating the  $n$-th order transformation. For instance, the first-order reduced gyro-angle $\bar\th$ is given by $\bfmg_1^\th$, which is gauge dependent. As for the first-order $\bar\sfc$, it is given by $(\bfmg_1^\bfq\cdott\na+\bfmg_1^\th~\p_\th)\sfc$, which is gauge-independent \cite{Little81, Little83, CaryBriz09}. 
 
 A second illustration is given by the guiding-center Poisson bracket. When using the coordinate $\th$, its expression is gauge dependent \cite{Little81, Little83} because of the presence of the gauge vector $\bfmr$. Nevertheless, it was noticed \cite{BrizHahm07} that this presence can be combined with the gradient through the combination $\na+\bfmr~\p_\th$. Actually, this is exactly the connection-independent gradient for the coordinate $\sfc$. Indeed, formally the covariant derivative $\bar\na$ can be written $\p_{\bfq|\sfc}+\bar\na\sfc\cdott\p_{\sfc|\bfq}$, which is connection-dependent, but in such a way as to make the quantity $\bar\na-\bar\na\sfc\cdott\p_{\sfc|\bfq}=\bar\na+\na\sfb\cdott\sfc~\sfb\cdott\p_\sfc - \bfmr_g \sfa\cdott\p_\sfc$ connection-independent. This last quantity is not a gradient because the second term in the right-hand side brings $\sfc$ out of its definition space. Thus, this term has to be removed to obtain the connection-independent gradient $\bar\na_*:=\bar\na + \bfmr_g ~\p_\th$. It is the minimal connection, since it formally writes $\bar\na_*=\p_{\bfq|\sfc}-\na\sfb\cdott\sfc~\sfb\cdott\p_{\sfc|\bfq}$, which corresponds to $\bar\na_*\sfc=-\na\sfb\cdott\sfc~\sfb$, or equivalently $(\bfmr_g)_*=0$. This minimal connection corresponds to the orthogonal projection, in the picture mentioned above where the connection is viewed as a projection from the circle $\mathbb S^1(\bfq)$ to the circle $\mathbb S^1(\bfq+\delta \bfq)$. 

 Also the content of the true intrinsic arbitrariness can have practical interest. For instance, the physical connection (\ref{ConnectNatur}) is not available when using the coordinate $\th$. It does not correspond to a gauge fixing (since it depends on the momentum variables). Neither does the connection $\bfmr=0$, which would simplify computations and remove the arbitrary terms from the theory. As a result, there is no natural gauge fixing in this case. It is why previous works tried to identify a natural gauge fixing based on the magnetic geometry \cite{Little88, Sugi08}. On the contrary, with the gauge-independent coordinate $\sfc$, the physical choice $\bfmr_g=-\cot\vph \na\sfb\cdott\sfa$ does correspond to a connection fixing. As for the simplifying choice $\bfmr_g=0$, it is just the minimal connection. It is also the natural geometrical connection, in the sense that it is induced by the definition of $\mathbb S^1(\bfq)$ and that it corresponds to the connection-independent gradient, as appeared in the previous paragraph.
 
 Thus, the intrinsic approach is not just an optional reformulation of the theory. It emphasizes the intrinsic properties underlying the gauge arbitrariness, which concern the connection of the covariant derivative $\bar\na$ rather than the coordinate system. It also makes available some relevant gradients that were inaccessible in the gauge-dependent formulation.

	\section{Intrinsic counterpart of the anholonomy}

 We now turn to the remnants of the traditional guiding-center anholonomy in the intrinsic approach. Anholonomy means that, after performing one closed path in some space, some partial contribution does not sum up to zero or some quantity defined in another space with a given parallel transport does not recover its initial value. The guiding-center anholonomy initially concerns the gauge $\sfe_1$ and can be considered from two complementary points of view \cite{Little81, Little88}. 
 
 In the first point of view, one performs a closed path $\ga$ in configuration space with $\sfe_1$ parallel transported along the connection $\bfmr$. At the end of the loop, the vector $\sfe_1$ recovers its value but the sum of its infinitesimal rotations around the magnetic field corresponds to a non-zero angle \cite{Little81, Little88}
\bq
	\Delta \th_g
	:=
	\oint_\ga (d\bfq\cdott\na\sfe_1)\cdott\sfb\x\sfe_1
	=
	\int_S \na\x\bfmr \cdott d\mathbf S
	\neq 0
	\,,
\label{AnholonScalarGauge}
\eq
where $S$ is a surface with boundary $\p S=\ga$. In the second point of view, the loop $\ga$ is performed while preventing $\sfe_1$ to rotate around the magnetic field (i.e. it is moved in such a way that $\delta\bfq\cdott\na\sfe_1\cdott\sfe_2=0$). Then at the end of the process, the vector $\sfe_1$ does not recover its value: it has rotated by a non-zero angle, which is given by $-\Delta\theta_g$.

 Notice that Eq.~(\ref{AnholonScalarGauge}) assumes that the loop $\ga$ is contractible. In this paper, we will always consider that it is the case (e.g. the space is simply connected). Otherwise $S$ does not exist and there is no relation equivalent to the second equality in (\ref{AnholonScalarGauge}). This simply-connected assumption is enough to study the local structure of the system, since locally in a three-dimensional domain any loop is contractible. In order to study also global aspects, the assumption must be released. For instance in a tokamak not all loops are contractible. This should not cause a problem because the results of the next section will agree with the results of Ref.~[\onlinecite{BurbQin12}], where the assumption on contractible loops is not used. Thus this assumption plays no essential role. It is useful only to simplify the argument. 

 The non-zero angle (\ref{AnholonScalarGauge}) impacts the coordinate $\theta$, whose variations do not depend only on the state of the particle but also on the gauge fixing $\sfe_1$. The partial contribution due to the gyro-gauge is called the \textsl{geometric phase}, denoted by $\Delta \th_g$. It is anholonomic because of Eq. (\ref{AnholonScalarGauge}). This anholonomy term cannot be made zero by a choice of gauge, because its integrand is given by \eqt{\ref{VectCurlGaugeVect}}, which is gauge-independent. This fact suggests that the anholonomic quantity (\ref{AnholonScalarGauge}) is related to an intrinsic property of the system. It is why the anholonomy was considered as unavoidable in guiding-center coordinates \cite{Little81, Little88, NoteAnholonomy}. \\

 With the gyro-angle $\sfc$, the anholonomy does not concern the coordinate system, since the coordinates are defined directly from the physical state. There is no extrinsic quantity (such as $\sfe_1$) implied in the definition of this gyro-angle to generate anholonomy. However, this does not preclude the possible presence of anholonomy in the intrinsic framework. To investigate this point, we will first identify the intrinsic counterparts of the quantities involved in the traditional guiding-center anholonomy, mainly the total variation of the gyro-phase $d\theta$ and the geometric phase $\Delta\theta_g$. Then the question of anholonomy in the gauge-independent framework will become clear. \\

 To begin with, the gauge-independent gyro-phase $\sfc$ is not a scalar angle. It is a vector, and its infinitesimal variations are two-dimensional, as shown in \eqt{\ref{ConnectGene}}. For an intrinsic description of the phenomenon at work in \eqt{\ref{AnholonScalarGauge}}, it is convenient to identify a scalar quantity for the variation of the gyro-angle $\sfc$. 
 
 The change of $\sfc$ in the direction $\sfb$ is not relevant, since it just corresponds to maintaining $\sfc$ in its definition domain through spatial displacement. Thus, the effective variation of the gyro-angle is only in the direction of $\sfa$. In addition, the true change of $\sfc$ is obtained after removing the contribution coming from the spatial displacement. From this point of view, 
\begin{equation}
	\delta \Theta
	:=-\sfa \cdott(d\sfc-d\bfq\cdott\bar\na\sfc)
	=-\sfa \cdott d\sfc+d\bfq\cdott\bfmr_g
\label{deltaTheta}
\end{equation}
is the quantity measuring the Larmor gyration. The minus sign in the prefactor is a convention in order to agree with the usual orientation for the gyro-angle $\th$. 
 
 The relevance of the scalar variation $\de\Theta$ is emphasized by the fact that the set of $1$-forms
\bq
	\Big(
		d\bfq, dp, d\varphi, \delta \Theta
	\Big)
\label{Basis1Forms}
\eq
is dual to the natural derivative operators of the theory 
\bq
	\Big(
		\bar\na, \p_p, \p_\varphi, \p_\th:=-\sfa\cdott\p_\sfc
	\Big)
		\,.
\eq
Here we insist that the generator of Larmor gyration is written $\p_\th$ but its definition does not depend on the gauge.

 In addition, when the local gauge-dependent description for the gyro-angle is used (implicitly chosen as usual such that inside the local description there is a globally defined $\sfe_1$ and $\bfmr_g$ is defined by $\bfmr$), it is readily checked that $\delta\Theta=d\th$. This observation confirms that $\delta\Theta$ is the intrinsic (global) quantity corresponding to $d\th$. As a corollary, it explains why the $1$-form $d\th$ is gauge dependent and the associated gauge-independent $1$-form is $d\th-d\bfq\cdott\bfmr$. Indeed, the reason is that $\de\Theta$ depends on $\bfmr_g$ and the associated connection-independent quantity is $-\sfa\cdott d\sfc=\de\Theta-d\bfq\cdott\bfmr_g$. An essential difference compared to $d\th$ is that $\delta\Theta$ is not closed:
\bq
	d(\de\Theta)
	= 
	-
	(d\bfq\cdott\na\sfb)\cdott(\sfb\x\sfb'd\bfq)
	+
	d\bfmr_g\cdott \wedge~ d\bfq
	\,,
\label{dDeltaTheta}
\eq
where for notational convenience, the primed notation is used for gradients acting on their left: $\sfb'd\bfq=d\bfq\cdott\na\sfb$. The wedge symbol $\wedge$ indicates antisymmetry: $a.\wedge~ b=a.b-b.a$. 

 In Eq.~(\ref{dDeltaTheta}), the magnetic term can be rewritten by using Eq.~(\ref{FormAgreeLittlejohn}), in agreement with Littlejohn's results \cite{Little88}. However, Littlejohn derived the curvature of $\delta\Theta$ in the special case where $\bfmr_g=0$. He did not consider a more general connection, nor did he indicate that there was any freedom in selecting the connection. The interest of Eq.~(\ref{dDeltaTheta}) is that it applies to any connection $\bfmr_g$. 
 
 For completeness, let us mention that, with the most general connection for the gyro-angle $\sfc$, Eq.~(\ref{deltaTheta}) is written as 
\bq
	\delta \Theta
	=
	-\sfa \cdott d\sfc+d\bfq\cdott\bfmr_g
	+ f_1 dp + f_2 d\varphi
\,,
\eq
where $f_1$ and $f_2$ are arbitrary phase-space functions. Then Eq.~(\ref{dDeltaTheta}) would become
\bq
	d(\de\Theta)
	= 
	-
	(d\bfq\cdott\na\sfb)\cdott(\sfb\x\sfb'd\bfq)
	+
	d\bfmr_g\cdott \wedge~ d\bfq
	+
	df_1 \wedge dp + df_2 \wedge d\varphi
	\,.
\eq
Non-zero $f_1$ and $f_2$ would imply that the connection and the covariant derivative concern not only spatial displacements (i.e. variations of the coordinate $\bfq$), but also variations of the two other non-gyro-angle coordinates $p$ and $\varphi$, as in Eq.~(\ref{ConnectGeneralized}). Since this refinement is useless here, for the following we will set $f_1=f_2=0$ and remain with only the connection $\bfmr_g$.\\

 The variation of $\Theta$ is defined along a path $\gamma$ by $\Delta\Theta=\int_\gamma \delta\Theta$. After performing one closed path that is the boundary of some surface $S$, $\Delta\Theta$ is given by: 
\bq
	\Delta\Theta
	=
	\oint_\gamma \de\Theta
	=
	\int_{S} d(\de\Theta)
	\neq 0
	\,,
\label{AnholonScalarIntrin}
\eq
which is non-zero in general. Thus, $\Theta$ is not a holonomic quantity. 

 More precisely, using \eqt{\ref{dDeltaTheta}}, the anholonomy (\ref{AnholonScalarIntrin}) of the scalar angle can be written
\bq
	\Delta \Theta
	=
	\Delta \Theta_B
	+
	\Delta \Theta_c
	\,.
\eq
The first contribution comes from the magnetic geometry  
\bq
	\Delta \Theta_B
	=
	-
	\int_{S}
	(d\bfq\cdott\na\sfb)\cdott(\sfb\x\sfb'd\bfq)
	\,.
\label{AnholonMagnetContrib}
\eq 
The second contribution comes from the choice of connection 
\bq
	\Delta \Theta_c
	=
	\int_{S} 
	d\bfmr_g\cdott\wedge~ d\bfq
	\,,
\label{AnholonConnexContrib}
\eq 
and it is expected to be the intrinsic counterpart of the geometric phase.\\

 When the local gauge-dependent coordinate $\th$ is used for the gyro-angle, the connection vector $\bfmr_g=\bfmr=\na\sfe_1\cdott\sfe_2$ depends only on $\bfq$. Then the integrand of the contribution (\ref{AnholonConnexContrib}) exactly compensates the magnetic contribution (\ref{AnholonMagnetContrib}). Indeed, it writes
\begin{align}
	d\bfmr\cdott\wedge~ d\bfq
	&=
	(d\bfq\cdott\na\sfe_2) \cdott\wedge~ (d\bfq\cdott\na\sfe_1)
\label{AnholonConnContribGaugeCase}
\\
	&
	=
	(d\bfq\cdott\na\sfb)
	\cdott(\sfe_2\sfe_1-\sfe_1\sfe_2)\cdott(\sfb' d\bfq)
\ncr
	&
	=
	(d\bfq\cdott\na\sfb)\cdott\sfb\x(\sfb' d\bfq)
	\,,
\notag
\end{align}
where the first equality comes from the antisymmetry, and the second comes by inserting the identity matrix $(\sfb\sfb+\sfe_1\sfe_1+\sfe_2\sfe_2)\cdott$ and by using the relations
\begin{align}
	(d\bfq\cdott\na\sfe_1)\cdott\sfe_1&=0
	\,,
	\ncr
	(d\bfq\cdott\na\sfe_2)\cdott\sfe_2&=0
	\,,
	\ncr
	(d\bfq\cdott\na\sfe_1)\cdott\sfb 
	&=-(d\bfq\cdott\na\sfb)\cdott\sfe_2
	\,,
	\notag
\end{align}
which come because $(\sfb,\sfe_1,\sfe_2)$ is an orthonormal basis. 

 For a comparison with the anholonomic phase (\ref{AnholonScalarGauge}), the integrand of the connection contribution (\ref{AnholonConnexContrib}) can be written by using that it is only a function of the position:
\begin{align}
	d\bfmr\cdott\wedge~ d\bfq
	&=
	d\bfq\cdott(\na\bfmr-\bfmr') ~d\bfq
\ncr
	&
	=-d\bfq\cdott(\na\x\bfmr)\x d\bfq
	\,,
\notag
\end{align}
which agrees with (\ref{AnholonScalarGauge}).

 These results can be shown to agree with Littlejohn's expression (\ref{VectCurlGaugeVect}) by using the antisymmetry of the matrix $\na\sfb\cdott\sfb\x\sfb'$ in order to write it as a cross product:
\begin{align}
		(d\bfq\cdott\na\sfb)\cdott(\sfb\x\sfb' d\bfq)
		&
		=
		d\bfq^i \na^i\sfb\cdott\sfb\x\na^j\sfb ~d\bfq^j
\label{FormAgreeLittlejohn}
\\
		&
		=
		d\bfq^i d\bfq^j 
		\tfrac 12 (\de^{ik}\de^{jl}-\de^{il}\de^{jk}) 
		\na_k\sfb\cdott\sfb\x\na_l\sfb 
	\ncr
		&
		=
		d\bfq^i d\bfq^j 
		\tfrac 12 \ep_{ijA}\ep_{Akl} 
		\na_k\sfb\cdott\sfb\x\na_l\sfb 
\ncr
		&
		=
		\ep_{iAj} d\bfq^i 
		\left(-\tfrac{1}{2}\right)
		\ep_{Akl} 
		\na_k\sfb\cdott\sfb\x\na_l\sfb 
		~d\bfq^j 
\ncr
		&
		=
		d\bfq\cdott\mathbf N \x d\bfq
		\,,
	\notag
\end{align} 
with
\begin{align}
	\mathbf N_A
	:
	=
	&
	\left(-\tfrac{1}{2}\right)
	\ep_{Akl} 
	\ep_{\alpha\beta\ga}
	\na_k\sfb_\alpha\sfb_\beta\na_l\sfb_\ga 
\ncr
	=
	&
	\tfrac{\sfb_A}{2}
	\big(
	\text{Tr}(\na\sfb\cdott\na\sfb)
	-
	(\na\cdott\sfb)^2
	\big)
\ncr
	& +(\na\cdott\sfb)~(\sfb\cdott\na\sfb)_A
	-
	(\sfb\cdott\na\sfb\cdott\na\sfb)_A
	\notag
	\,.
\end{align} 
This is exactly Littlejohn's expression (\ref{VectCurlGaugeVect}). In the computation (\ref{FormAgreeLittlejohn}), the first equality comes from the antisymmetry of the matrix $\na\sfb\cdott\sfb\x\sfb'$, and all the other equalities are properties of the Levi-Civit\'a symbol. An alternative (more direct but heavier) way to prove the result is to insert the identity matrix $(\sfb\sfb+\sfe_1\sfe_1+\sfe_2\sfe_2)\cdott$ everywhere in $(d\bfq\cdott\na\sfb)\cdott(\sfb\x\sfb' d\bfq)$. Then expanding and simplifying the formula gives the expected result. 

 So, in the local gauge-dependent case, the connection contribution is exactly the geometric phase
\bq
	\Delta \Theta_c 
	=
	\Delta \theta_g
	\,,
\eq
which confirms that $\Delta \Theta_c$ is the intrinsic counterpart of the geometric phase. In addition, Eq.~(\ref{AnholonConnContribGaugeCase}) shows that, when the coordinate $\th$ is used, the gauge contribution is exactly the inverse of the anholonomic magnetic contribution (\ref{AnholonMagnetContrib}):
\bq
	\Delta \Theta_c 
	=
	-\Delta \Theta_B
	\,.
\eq
This equality explains both that the corresponding gyro-angle $\th$ is holonomic $d(\delta\Theta)= d^2\th=0$, and that the associated geometric phase $\Delta\th_g$ is anholonomic.\\

 A first consequence can be noticed by now: in the intrinsic approach, the counterpart of the geometric phase is arbitrary and it can be made holonomic. Even its integrand can be set to zero, since  $\bfmr_g$ can be chosen freely. So, the anholonomic geometric phase observed in the gauge-dependent approach is not intrinsic in itself. On another hand, choosing the geometric phase zero makes the total phase $\Delta \Theta$ anholonomic, since in that case it is exactly given by the anholonomic term (\ref{AnholonMagnetContrib}) due to the magnetic geometry. This suggests that in the intrinsic framework anholonomy is not only present but inherent to the structure of the system.\\

 To investigate this point, let us emphasize that the introduction of a scalar variation for $\sfc$ was used above only to identify the correspondence between the local and global descriptions. From an intrinsic point of view, what we have obtained is just that $\de\Theta$ is not closed. Hence $\Theta$ is not a proper coordinate, but it is not needed since the gyro-angle coordinate is the vectorial quantity $\sfc$. All the same, anholonomy effects can be viewed even in this framework from two complementary (dual) points of view.

 The first of them considers the properties of the basic $1$-forms (\ref{Basis1Forms}) of the theory. The previous investigations showed that the basic differential form for the gyro-angle is $\delta \Theta$, and that it is not closed. This implies anholonomy, as appeared in Eq.~(\ref{AnholonScalarIntrin}) and as can be found in textbooks, e.g. in Ref.~[\onlinecite{MarsdenMontRati}]. In the fiber-bundle approach, this conclusion is still clearer, since the anholonomy is given by the curvature $2$-form \cite{NoteCurvature}, here $d(\delta\Theta)$. 
  
 The second point of view more basically considers the properties of the elementary vector fields of the theory. Indeed, the effects of an infinitesimal loop in configuration space (with the parallel transport defined by the connection) are evaluated with the commutator of gradients, as is confirmed in Ref.~[\onlinecite{HamilRedStag}] or~[\onlinecite{Lang95}] for instance. Here, the commutator is 
\bq
	[\bar\na_i,\bar\na_j] 
	=
	\Big(
		\na_i\sfb\cdott\sfb\x\na_j\sfb
		-
		\bar\na_{i}(\bfmr_g)_{j}
		+
		\bar\na_{j}(\bfmr_g)_{i}
	\Big)
	~\p_\th
	\,,
\label{CommutGrad}
\eq
which is dual to \eqt{\ref{dDeltaTheta}}. 

 The non-commutation of gradients in \eqt{\ref{CommutGrad}} is a consequence of the presence of a constrained coordinate system, with its associated non-zero connection. It means that the action of gradients does not fit directly with the coordinate system. After a closed path in the sense of the gradients, i.e. of the sum of infinitesimal variations, the coordinates do not recover their initial value. Conversely, after one loop in coordinate space, the coordinates recover their initial value, but the sum of infinitesimal changes is not zero. Thus, not only is anholonomy inherent to the introduction of a scalar angle, which generalizes the conclusion of Refs.~[\onlinecite{Little81, Little88}], but it is an intrinsic feature of the space of particle states $(\bfq, p, \vph, \sfc)$. This feature, although absent from trivial coordinate systems, is not an issue. It is quite common in spaces with non-zero curvature, e.g. in general relativity. 

 The questions come with the gauge-dependent approach when requiring a coordinate system that fits with the action of gradients. This is possible only when the geometry of the bundle is trivial. In addition, in the resulting trivialized space the unavoidable presence of non-trivial anholonomy effects become puzzling. So, the scalar coordinate $\th$ for the gyro-angle has holonomic (commuting) gradients but all the same involves anholonomic (unphysical) phases. In addition, it is valid only locally because it loses (makes trivial) the geometry of the coordinate space. On the contrary, the global gyro-angle $\sfc$ has anholonomic gradients rather than anholonomic phases, but it retains all the geometry of the guiding-center coordinate system. 

 This necessary alternative comes from \eqt{\ref{dDeltaTheta}} or (\ref{CommutGrad}), which shows that anholonomy is unavoidable and intrinsically related to the magnetic geometry. The anholonomic term $\na\sfb\cdott\sfb\x\sfb'$ has to be put either as a non-zero commutator of gradients or as an anholonomic phase for the gyro-angle.

\section{Towards a scalar intrinsic gyro-angle}

 The previous three sections were concerned with the first goal of this paper. They showed how the intrinsic approach clarifies the questions raised by the presence of a gyro-gauge. They also emphasized the true intrinsic properties that were underlying in the traditional guiding-center anholonomy and gauge arbitrariness. We now turn to the second goal of this paper, which is to investigate how the intricacies caused by the presence of anholonomy and of covariant derivatives can be eliminated from the intrinsic approach. The former is studied in this section, while the latter will be considered in the next section. 
 
 So, we are presently interested in removing the anholonomy effects observed in the previous section. The basic idea is to use the freedom embodied in the connection in order to make the commutator of gradients zero. This question comes very timely since one point is to be clarified about our previous results. When the connection is given its physical expression (\ref{ConnectNatur}), the gyro-angle $\sfc$ is just the perpendicular velocity. It should be holonomic (i.e. it should not have non-zero commutators of gradients), since it is directly given by the physical momentum and the magnetic field, both of which are holonomic. \\

 The reason for this anholonomy is that the connection was defined through the physical definition of $\sfc$, but not the physical definition of the whole momentum. In order to take into account the whole momentum, a more general connection should be used, affecting also the pitch-angle $\vph$. 
 
 The variable $\vph$ is not a constrained coordinate, since it is defined over an independent space $\mathbb R^1$. Unlike the variable $\sfc$, it does not have to change value through spatial transportation, but it is allowed to. A flat (zero) connection is possible but it is only the trivial choice, analogous to the choice $\bfmr_g=0$ for the coordinate $\sfc$. In the same way as the free term $\bfmr_g$, the coordinate $\varphi$ can have an arbitrary connection. Especially, its definition from the physical momentum through \eqt{\ref{GyroangleIntrDef}} induces a non-zero connection
\bq
	\p_{\bfq|\bfp} \vph =-\na\sfb\cdott\sfc 
	\,.
\label{ConnectPhi}
\eq
Notice that in this argument two different gradients are implied: the one in the initial coordinates $\na=\p_{\bfq|\bfp}$ and the one in the final coordinates $\bar\na$. This last is roughly $\p_{\bfq|p,\vph,\sfc}$ but it takes into account the necessary connection for $\sfc$ and the possible connection for $\vph$. When acting on functions of $\bfq$ only, e.g. in the right-hand side of \eqt{\ref{ConnectPhi}}, they are equal, but in general they are not. What we call the \textsl{physical connection} is the one that makes them equal. For instance, the action of the associated covariant derivative on the quantity $\vph$ is defined by
\bq
	\bar\na\vph=\p_{\bfq|\bfp} \vph
	\,.
\eq
Notice also that within the constrained-coordinate picture the presence of a non-zero connection for $\varphi$ causes no complication. In the fiber-bundle picture, it would imply to change the framework, because the coordinate $\varphi$ would have to be considered in the fiber, not in the base space. \\

 The physical relevance of this connection can be viewed in the components $\mathbf V_i$ of the velocity vector field. They are defined by the relation $\dot f = \mathbf V_i\cdott \p_i f$ for any function $f$ of the phase space. Because of the non-trivial connection, they are different from the components of the velocity $\dot \bfz_i$, where the vector $\bfz=(\bfq,p,\vph,\sfc)$ combines all the coordinates. The relation between $\dot \bfz_i$ and  $V_i$ is given by
\bq
	\dot\bfz_i
	=
	\tfrac{d}{dt}\bfz_i
	=
	\sum_j\mathbf V_j\cdott \p_j \bfz_i
	=
	\mathbf V_i + \mathbf K_i
	\,,
\label{VelocityVectFieldCfVelocity}
\eq
where $\mathbf K_i:=\sum_j (1-\de_{ij}) \mathbf V_j\cdott \p_j \bfz_i$ is a connection term for it does not contribute when $\p_j\bfz_i= 0$ for $i\neq j$. In the specific case we are considering, the connection is involved only when a gradient acts on the coordinates $\sfc$ or $\varphi$. So, \eqt{\ref{VelocityVectFieldCfVelocity}} can be interpreted as
\bq
		\tfrac{d}{dt} \sfc = (\p_t + \dot\bfq \cdott \bar\na ) \sfc
	\,,
\eq
together with the same formula with $\sfc$ replaced by $\varphi$.

 With the physical connections (\ref{ConnectPhi}) and (\ref{ConnectNatur}) for the pitch angle and the gyro-angle, the components of the velocity vector field are given by
$$
	\mathbf V_i
	:=
			\left(
				\begin{smallmatrix}
				\tfrac{\bfp}{m} \\ 
				\tfrac{e\bfme\cdott\bfp}{p} \\ 
				\tfrac{e\bfme}{p\sin\vph}
				\cdott\left(\cos \vph ~\tfrac{\bfp}{p} 
									- \sfb\right) \\ 
				- \tfrac{eB}{m}\sfa
				+ \tfrac{e \bfme\cdott \sfa}{p ~ \sin\vph}\sfa
				\end{smallmatrix}
			\right)
	\,.
$$
They perfectly agree with the physical force, which is just the Lorentz force. Especially, the limit where there is no electric field $\bfme=0$ is expressive: there remain only the velocity $\mathbf V_\bfq=\bfp/m$ and the Larmor gyration $\mathbf V_\sfc=-eB\sfa/m$. All the additional terms in the components $\vph$ and $\sfc$ of equations (\ref{Dynamics}), which do not come from the physical dynamics but from the magnetic geometry, are absorbed in the connection. This is satisfactory since the role of the connection is precisely to encode the change of the momentum coordinates through spatial displacement as a result of the magnetic geometry. 

 When using the coordinate $\th$, the geometric contributions in \eqt{\ref{DynamicsTheta}} cannot be absorbed in a connection contribution. In fact, the scalar coordinate $\th$ is precisely introduced to make the coordinate system trivial, and hence to have flat connection. Providing $\th$ with the corresponding connection would amount to using the intrinsic approach, with an additional detour by the gauge $\sfe_1$. 
 
 The dynamics of the gyro-angle can be reinterpreted in this light, in relation with Refs.~[\onlinecite{LiuQin11, BrizGuil12}]. In the dynamics (\ref{DynamicsTheta}) of the gyro-angle $\th$, only the first and fourth terms are contributions due to the physical dynamics. The second term corresponds to the so-called "adiabatic phase" in the case considered by Ref.~[\onlinecite{LiuQin11}]. This term comes from the magnetic term in the physical connection (\ref{ConnectNatur}), related to the definition for $\sfc$ to be physically the unit vector of the perpendicular momentum. It is induced by the change of the projection as a result of the change of the magnetic field (through spatial displacement). Thus it concerns also the intrinsic gyro-angle $\sfc$, e.g. in \eqt{\ref{Dynamics}} or \eqt{\ref{ConnectNatur}}. In addition, it is expected to be adiabatic only for the specific case considered by Ref.~[\onlinecite{LiuQin11}], but not for a general (inhomogeneous) strong magnetic field. This is confirmed by Ref.~[\onlinecite{BrizGuil12}]. As for the third term in \eqt{\ref{DynamicsTheta}}, it is the "geometric phase" and it is actually a gauge phase since it is purely related to the choice of gauge. It is absent from the intrinsic dynamics (\ref{Dynamics}) or connection (\ref{ConnectNatur}). \\

 With the full physical connection given by (\ref{ConnectNatur}) and (\ref{ConnectPhi}), although the coordinates $\sfc$ and $\vph$ are not independent of the variable $\bfq$, they behave exactly as the components of a vector $\hv:=\sfb\cos\vph+\sfc\sin\vph$ that is independent of $\bfq$, i.e. it has flat connection: 
\begin{align}
	\bar\na\hv
	&
	=\bar\na(\cos\varphi\sfb+\sin\varphi\sfc)
	\ncr
	&
	=\cos\varphi \na\sfb + \sin\varphi~\bar\na\sfc
	+\bar\na\varphi~(-\sfb\sin\varphi+\sfc\cos\varphi)
	=0
	\notag
	\,.
\end{align}
Actually, this computation shows that the flat connection $\bar\na\hv=0$ is obtained if and only if the connection is the full physical one. The vector $\hv$ stands for the unit vector of the momentum 
\bq
	\hv:=\tfrac{\bfp}{p}
	\,.
\label{DefinitVarHv}
\eq 

 As a consequence and as expected from physical intuition, the commutator of gradients with this connection is zero:
\bq
	[\bar\na_i,\bar\na_j]=0
	\,,
\eq
as is easily verified by direct computation. It traduces that, after one loop in configuration space (with $\bfp$ constant), both the momentum and the magnetic field come back to their initial value. \\

 The full physical connection given by (\ref{ConnectNatur}) and (\ref{ConnectPhi}) achieves a part of our goal by making the commutators of gradients zero, but other commutators have to be considered as well. Indeed, in the gauge-independent approach the space defined by the bundle is not the space of all $(\bfq,\sfc)$ but the whole phase space $(\bfq, p, \vph, \sfc)$. So, all the above conclusions apply after replacing the gradients $\bar\na$ by the complete set of basic derivative operators $\p_\bfz=(\bar\na, \p_p, \p_\varphi, \p_\th)$. Between these operators, even with the physical connection, the non-triviality of the fiber bundle for a general magnetic geometry should imply non-zero commutators. This is confirmed in \eqt{\ref{CommutGeneral}}, where non-trivial commutators of the basic derivative operators are given for a general connection, both for the gyro-angle $\bfmr_g:=\bar\na\sfc\cdott\sfa$ and for the pitch-angle $\bfmr_\vph:=\bar\na\varphi$. This includes all of the four choices of connection previously mentioned as special cases: the connection (\ref{ConnectGauge}) for the gauge-dependent case with coordinate $\th$; the physical connection (\ref{ConnectNatur}) for $\sfc$; the general connection (\ref{ConnectGene}) for $\sfc$; and the full physical connection (\ref{ConnectNatur}) and (\ref{ConnectPhi}) for $\sfc$ and $\vph$. 
\begin{align}
	\lbrack\bar\na_i,\bar\na_j\rbrack 
		&=~ 	
			\na_i\sfb\cdott\sfb\x\na_j\sfb ~\p_\th
\ncr
			&~~~~~
			- 	\Big( 
					\bar\na_i(\bfmr_g)_j - \bar\na_j(\bfmr_g)_i 
				\Big) ~\p_\th
\ncr
			&~~~~~
			+ 	\Big( 
					\bar\na_i(\bfmr_\vph)_j - \bar\na_j(\bfmr_\vph)_i 
				\Big) ~\p_\vph
			\,,
\label{CommutGeneral}
\\
	\lbrack\p_p,\bar\na\rbrack 
		&=~ 
			-\p_p\bfmr_g~\p_\th+\p_p\bfmr_\vph~\p_\vph
			\,,
\ncr
	\lbrack\p_\vph,\bar\na\rbrack 
		&=~ 
			-\p_\vph\bfmr_g~\p_\th
			+\p_\vph\bfmr_\vph~\p_\vph
			\,,
\ncr
	\lbrack\p_\th,\bar\na\rbrack 
		&=~
			 -\p_\th\bfmr_g~\p_\th
			 +\p_\th\bfmr_\vph~\p_\vph
			\,.
\notag
\end{align}

 In practical cases the second commutator in Eq.~(\ref{CommutGeneral}) is zero because $\bfmr_g$ does not depend on $p$. The reason is that the gyro-angle comes from the splitting of the coordinate $\hv$ into the pitch-angle and the gyro-angle \textsl{via} the magnetic geometry $\bfmb(\bfq)$. The coordinate $p$ plays no role in the process.

 \eqt{\ref{CommutGeneral}} clearly emphasises the crucial role of the anholonomic magnetic term $\na\sfb\cdott\sfb\x\sfb'$, which is the only affine term in the connection. For the minimal connection this magnetic term is the only non-zero term, which indeed simplifies computations. As for the full physical connection, it cancels the magnetic term, and also the whole commutator of gradients. However, the two commutators $\lbrack\p_\vph,\bar\na\rbrack$ and $\lbrack\p_\th,\bar\na\rbrack$ become non-zero. For instance $\p_\vph\bfmr_g\neq 0$ and $\p_\th\bfmr_\vph\neq 0$. \\

 With the general setting considered in \eqt{\ref{CommutGeneral}}, one can look for a connection that would make all commutators zero. This would provide a splitting of the vector $\hv$ into proper scalar coordinates for the pitch-angle and the gyro-angle, i.e. coordinates that fit with the action of commuting derivative operators. These coordinates would be defined from the value of the quantities $\vph$ and $\sfc$ at one point in phase space through parallel transportation by the commuting derivative operators \cite{NoteMonodromy}.
 
 A solution is not expected to be generally possible since a scalar coordinate for the gyro-angle means that the circle bundle is trivial. The goal is to identify the existence condition for the desired coordinate system. Indeed, the free $4$-dimensional function of phase space $\big(\bfmr_g(\bfz),\bfmr_\vph(\bfz)\big)$ opens new possibilities. One can consider using this larger freedom to obtain more complete results than with the gauge-dependent framework, whose freedom corresponds only to the $1$-dimensional gauge function of position space $\psi(\bfq)$ in \eqt{\ref{GaugeVectGaugeDepend}}. 

 The last three rows in \eqt{\ref{CommutGeneral}} imply that a solution $(\bfmr_g, \bfmr_\vph)$ must not depend on $\vph$, nor $p$, nor $\sfc$, hence it must be purely position-dependent. In addition, because of the third row in \eqt{\ref{CommutGeneral}}, $\bfmr_\vph$ must be curl-free. Actually, $\bfmr_\vph$ is useless and can be set to zero. As for the first two rows, they imply that $\bfmr_g$ must cancel the anholonomy term, which is purely position-dependent. Thus, the equation for the desired connection is  
\bq
	0=
	\na_i\sfb\cdott\sfb\x\na_j\sfb 
			- 	\bar\na_i(\bfmr_g)_j 
			+ \bar\na_j(\bfmr_g)_i 
				\,,
\eq
which can be rewritten
\bq
	\bar\na\x\bfmr_g
	=
	\mathbf N
	\,.
\label{CondCoordScal}
\eq

 This equation is reminiscent of the usual relation (\ref{ConditionIntegrGaugeVect}) in the gauge-dependent approach, but they are different both in their origin and in their meaning. On the one hand, \eqt{\ref{CondCoordScal}} is obtained without appealing to the idea of a gyro-gauge nor its associated gyro-angle $\theta$. Starting from the structure of the manifold defined by the physical coordinates $(\bfq, p, \vph, \sfc)$, we are looking for an arbitrary scalar gyro-angle coordinate. The solution might not be related to a choice of gauge. More precisely we are looking for a connection corresponding to this coordinate. On the other hand, Eq.~(\ref{CondCoordScal}) is a necessary and sufficient condition for the existence of a scalar gyro-angle, whereas in previous works, the analogous relation (\ref{ConditionIntegrGaugeVect}) only appeared as a consequence of the existence of a gauge. 

 When a solution $\bfmr_g:=\bfmr_s$ of \eqt{\ref{CondCoordScal}} exists, the scalar gyro-angle coordinate is defined from parallel transportation with the derivative operators defined by the associated connection $\bfmr_s$. This parallel transportation results in a (trivializing) global section of the circle bundle, which in turn provides a zero for measuring the gyro-angle. Now, the important point in the above analysis is that $\bfmr_s$ depends only on the position. The parallel transportation actually results in a section of the restricted circle bundle over the position space. This property means that a scalar coordinate always defines a gyro-gauge. It is the reciprocal of the property that a gyro-gauge provides a scalar gyro-angle, which was considered in previous works and in the previous sections. 
 
 As a consequence, condition (\ref{CondCoordScal}) is also a necessary and sufficient condition for a global gauge to exist. Here, it is obtained in a direct argument on commuting derivatives (but under the assumption of contractible loops). This is very different from the work by Burby and Qin where the existence of a global gauge was studied \cite{BurbQin12}: the proof for the necessary condition used an auxiliary property and the proof for sufficiency required "a lengthy digression into the theory of principal bundles and characteristic classes". Finally, the Burby-Qin condition for the existence of a gauge is slightly different from ours, but they are equivalent. Their condition states that through the boundary $S$ of any hole inside the spatial domain the vector field $\mathbf N$ has zero net flux:
\bq
	\oint_S \mathbf N\cdott d\mathbf S =0
	\,,
\eq
and this is the boundary condition for the solvability of \eqt{\ref{CondCoordScal}}, since $\na\cdott\mathbf N=0$.

 Our derivation assumed contractible loops. It does not apply when the base space contains non-contractible loops, e.g. in a tokamak geometry. In this case, the work of Burby and Qin is needed to conclude about the existence of a gauge. In turn, the existence of a scalar angle is implied by the presence of a gauge. Since the existence condition is the same in any case and regards only non-contractible spheres, the assumption on non-contractible loops plays no essential role and can be released. A complete understanding of the origin of this fact requires a more detailed study, which is outside the scope of the present paper.

	\section{An intrinsic formalism with no covariant derivative}

 We now turn to the last aspect of the second goal of the paper, which is to simplify the formalism by removing the presence of covariant derivatives from the intrinsic coordinate system. 

 Indeed, Secs.~II-IV showed that the questions of the standard coordinate $\th$ were related to properties of the basic derivative operators involved in guiding-center theory, which do not fit with a standard coordinate system. In this sense, the philosophy of the gauge-independent approach is to consider separately the coordinate system and the basic derivative operators. It avoids putting in the coordinate system some properties that actually concern the derivative operators. 

 However, the resulting formalism may seem more complicated than expected, because of the constrained coordinate system, with the associated subtleties about covariant derivatives, non-zero commutators, non-closed basis of $1$-forms, etc. One can consider going one step further in the development of this intrinsic approach and removing these subtleties from the coordinate system. In one way or another, they are unavoidable in the theory since they result from properties of the non-trivial circle bundle implied by the fast guiding-center coordinate. But they concern the derivative operators, not the coordinate system, so that one can consider making the coordinate system both gauge-independent and unconstrained. To do so, it is natural to try to identify a scalar gyro-angle coordinate, as in the previous section. In such a coordinate system, covariant derivatives would automatically disappear. However, Sec.~V showed that a global scalar coordinate for the gyro-angle does not exist in general. Thus, the removal of covariant derivatives must be looked for by other means. \\

 Since the constrained coordinate system comes from the gyro-angle coordinate with its $\mathbb S^1$ fiber-bundle, a possible idea is again to come back to more primitive coordinates and to avoid the splitting of the momentum into the pitch-angle and the gyro-angle. Then, the unit vector of the momentum $\hv:=\tfrac{\bfp}{p}$ is kept as a single two-dimensional coordinate, as was approached by the results of the previous section. 
 
 Separating the pitch-angle from the gyro-angle is necessary at the end of the guiding-center reduction, when the gyro-angle is removed from the dynamics to obtain the slow reduced dynamics $(\dot{\bar\bfq},\dot{\bar\varphi})$. But at that point the fiber-bundle and the constrained coordinate $\sfc$ are also removed. In the course of the reduction process the splitting is not needed. What is needed is only a basis of $1$-forms and a basis of derivative operators that fit with the separation of scales: for instance to decompose the transformation of the vector $\hv$ between the contribution for the fast gyro-angle and the one for the slow pitch-angle; or to decompose the change of spatial coordinate into its components transverse and parallel to the magnetic field. 
 
 As a result, the method of using intrinsic coordinates and defining derivative operators adapted to the purpose can be applied. The interesting point is that the definition space for the coordinate $\hv$ is the sphere $\mathbb S^2$, whose immersion in $\mathbb R^3$ is independent of the spatial position. So, a trivial connection available, and this minimal connection is also the physical one since the definition (\ref{DefinitVarHv}) for $\hv$ does not depend on the position. 

 Notice that, since this coordinate is a two-dimensional vector immersed in $\mathbb R^3$, its variations are constrained. The operator $\p_\hv$ and the $1$-form $d\hv$ are purely transverse: $\hv\cdott\p_\hv=\hv\cdott d\hv=0$. But the coordinate system is not constrained any more: the coordinates are independent of each other. The basic differential operators  $(\p_\bfq, \p_p, \p_\hv)$ and $1$-forms $(d\bfq, dp, d\hv)$ behave trivially (i.e. practical calculations are similar to those using standard coordinates). \\

 For the purpose of the guiding-center reduction, the splitting between the pitch-angle and the gyro-angle is implemented in the basis of vector fields and 1-forms: $\p_\hv$ and $d\hv$ are decomposed to distinguish their contributions in the azimuthal direction $\sfa$ (corresponding to the variable $\th$) and in the elevation direction $\sfa\x\hv$ (corresponding to the variable $\varphi$). For instance, the operator $\p_\hv$ can be decomposed as $\left(\hv\x\sfb\cdott\p_\hv, \tfrac{(\sfb\x\hv)\x\hv}{\sin\varphi} \cdott\p_\hv\right)$, in order to agree with the traditional operators $(\p_\th, \p_\vph)$. Alternatively, the second operator can be chosen as $-\sin\varphi ~(\sfb\x\hv)\x\hv \cdott\p_\hv=\p_\phi$, in order to fit with the variable $\phi:=\cot\varphi$ that made formulae polynomials \cite{GuilMagMom, GuilGCmin, GuilMoyGa}. In a simpler way, it can be chosen just $(\sfb\x\hv)\x\hv \cdott\p_\hv$. This arbitrariness in the choice of a basis for vector fields is similar to the connection freedom in previous sections, but it is different. It concerns the splitting of the operator $\p_\hv$ rather than the definition of the gradient operator.
 
 This splitting procedure is only a generalization of what is commonly done for the position space. The three-dimensional quantity $\bfq$ is usually kept as a coordinate but the gradient $\na$ and the differential form $d\bfq$ are split into scalar components suited to the derivation, namely their components parallel to $\sfa$, $\sfb$, and $\sfc$. 

 It is straightforward to verify that guiding-center reductions work as usual within this formalism. In a similar way as what occurred when going from the coordinate $\th$ to the coordinate $\sfc$, the introduction of the coordinate $\hv$ removes all the intricacies from the coordinate system and confines them to the basis of vector fields of the theory. The only difference is that here the basis is not induced by the coordinate system, whose associated derivative operators are now trivial. It is induced by the purposes of the guiding-center reduction, such as the separation of scales. All the subtleties mentioned in the previous sections remain present but they are encoded in the properties of the chosen basis. For instance, when the basis is chosen with $\left(\hv\x\sfb\cdott\p_\hv, \tfrac{(\sfb\x\hv)\x\hv}{\sin\varphi} \cdott\p_\hv\right)$, the formalism is the same as in the previous section with the "full physical connection" (\ref{ConnectNatur}) and (\ref{ConnectPhi}).\\

 One could consider going one step further and keeping all the momentum coordinates $(p,\hv)$ as a single coordinate $\bfp$. This is unsure to be relevant, since the coordinate $p$ actually plays no role in the introduction of the gyro-angle, as we mentioned previously. 
 
 In addition, keeping the coordinate $\bfp$ does not seem convenient for practical computations. When the norm of the momentum $p$ is kept in the reduced coordinates, as in Ref.~[\onlinecite{GuilGCmin}], it is unchanged by the guiding-center transformation. Then it is useless to combine it with $\hv$, which is changed by the transformation. On another hand, most often the coordinate $p$ is replaced by the constant of motion conjugated to the gyro-angle, the magnetic moment $\bar\mu$. Then the coordinate $p$ is usually changed to $\mu$ in a preliminary step, in order for the remaining transformation to be near-identity. In this case, the splitting of the coordinate $\bfp$ into $p$ and $\hv$ is essential. 
 
 Last, separating the variables $p$ and $\hv$ is interesting for dimensional reasons. It implies that only one of the momentum coordinates is non-dimensionless, which can be very convenient for the derivation of the guiding-center reduction \cite{GuilMoyGa}. 
 
 These considerations do not mean that using the coordinate $\bfp$ is to be excluded. For instance, the recent work [\onlinecite{BurbySquir13}] proposed an algorithm for guiding-center reductions where the coordinate $\bfp$ was kept in a first near-identity transformation. Then the magnetic moment was identified and it could be adopted as a coordinate in a second transformation, not near-identity.

	\section{Conclusion}

 The gauge-independent approach of guiding-center theory clarifies the questions associated with the usual gyro-angle coordinate. The use of the physical gyro-angle as a coordinate removes the non-global existence, the gauge dependence and the anholonomy from the coordinate system. The corresponding coordinate system agrees both with the physical description and with the mathematical structure of the system, a non-trivial circle bundle. 

 This physical gyro-angle is constrained and position-dependent, which implies the presence of a covariant derivative encoding the geometry of the bundle. The induced connection involves a freedom, which is the intrinsic counterpart of the gauge arbitrariness but is much larger and very different. Especially, this freedom does not affect the coordinate system and embodies a choice in the basis of vector fields of the theory. 
 
 Because of the larger freedom, relevant choices become available. For instance, the connection can be chosen so as to fit with the physical definition of the gyro-angle. Alternatively, it can be set to zero. This minimal connection simplifies computations and removes the arbitrary terms from the theory. It was found to be underlying in previous results and can be considered as the natural geometrical connection. Each of these two choices does not correspond to a gauge fixing but to a connection fixing, which shows that the intrinsic formulation is needed for the description to fit with the physics and the mathematics of the guiding-center system. This is also emphasized by the fact that the question about non-global existence not only disappears but also has no counterparts in the intrinsic formulation. 
 
 Both the physical and the minimal covariant derivatives have non-zero commutators, which are the counterparts of the anholonomy of the gauge-dependent approach. Again, they do not concern the coordinate system but the basic derivative operators. A third choice of connection was identified, by giving to the pitch-angle the connection induced by its definition from the physical momentum. Then, covariant derivatives $\bar\na$ do commute, but other non-zero commutators appear in phase space, which traduce the non-triviality of the circle bundle defined by the gyro-angle. 
 
 In this framework, existence conditions for a splitting of the momentum into scalar coordinates for the pitch-angle and for the gyro-angle can be studied. The perspective is broader and complementary compared to the study of the existence of a gyro-gauge. The resulting condition is just the invertibility of a curl on a divergenceless vector field. It corresponds to boundary conditions on this vector field, in agreement with previous results. In this paper, the analysis used contractible loops, but this restriction appeared as optional. A detailed investigation of this point could bring additional information on the structure of the guiding-center system. \\

 As a result, the gauge-independent formulation exhibits the intrinsic properties underlying the questions about the standard gyro-angle. The formulation mainly replaces puzzling aspects of the coordinate system by non-trivial, yet regular, properties of the basic derivative operators. 

 The questions originated from the requirement for the coordinates and the basic derivative operators to behave trivially. This requirement can be obtained only when the circle bundle is trivial, i.e. only locally for a general magnetic geometry. In addition, it generates both the gauge dependence and the anholonomy question in the coordinate system. To agree with the physical system, the basic derivative operators must have non commuting properties. In turn, it means that there do not exist trivializing coordinates and that a constrained coordinate has to be used for the gyro-angle. 
 
 An idea underlying the intrinsic approach is that perturbation theory needs adapted derivative operators but not necessarily adapted coordinates. So, the coordinate system can be chosen as it is intrinsically, i.e. as it comes primitively. This idea can be generalized to avoid introducing the gyro-angle coordinate, with the intricacies caused by the constrained coordinate system. The gyro-angle and the pitch-angle can be kept as the single initial coordinate, which is the unit vector of the momentum. Then the $\mathbb S^1$-bundle is replaced by an $\mathbb S^2$-bundle. The corresponding structure is trivial and the formalism becomes elementary, i.e. with no covariant derivatives (trivial connection) nor any non-zero commutator. The resulting coordinate system is both intrinsic and unconstrained. For guiding-center perturbation theory, an adapted basis of derivative operators is defined, which encodes all the intrinsic properties of the circle bundle associated to the gyro-angle.

\vskip-6ex
~
	\section*{Acknowledgement}

\vskip-1ex
\noindent 
We acknowledge financial support from the Agence Nationale de la Recherche (ANR GYPSI). This work was also supported by the European Community under the contract of Association between EURATOM, CEA, and the French Research Federation for fusion study. The views and opinions expressed herein do not necessarily reflect those of the European Commission. The authors acknowledge fruitful discussions with Alain Brizard and with the \'Equipe de Dynamique Nonlin\'eaire of the Centre de Physique Th\'eorique of Marseille. They also thank Alain Brizard for suggesting many improvements in the English of the manuscript and the reviewer for his comments and suggestions, which were very helpful to improve the manuscript.


\begin{thebibliography}{78}
\bibitem{SingularityGC}
For instance, at order $1$, Eq.~(30c) in Ref.~[\onlinecite{Little83}] shows that $\bar \theta$ is singular in $p\sin\varphi=0$, as is also confirmed by Ref.~[\onlinecite{CaryBriz09}]. In addition, at order $2$, Eq.~(23) in Ref.~[\onlinecite{GuilGCmin}] shows that $\bar\varphi$ is also singular in  $p\sin\varphi=0$. At higher orders, this singularity is expected to affect all the coordinates, because it is related to the order in the cotangent of the pitch-angle compared to the order in the Larmor radius, as is emphasized in Refs.~[\onlinecite{GuilMagMom,GuilGCmin,GuilMoyGa}].
\bibitem{NoteAnholonomy}
In principle, anholonomy in a principal fiber bundle is not related to a coordinate system, but induced by a choice of connection. However, in the standard description of the guiding-center system, the fiber bundle is induced by $(\bfq, \sfe_1)$, and the choice of a connection on this bundle is implied in the choice of a gyro-gauge, which in turn corresponds to the choice of a coordinate $\theta$. It is why there is a direct correspondence between the anholonomy and the standard gyro-angle coordinate system, although the anholonomy does not concern $\theta$ itself, but the vector $\sfe_1$ through the parallel transport defined by the connection. 

Interestingly, when using the intrinsic gyro-angle $\sfc$, this correspondence disappears. As it should, the anholonomy is not related to the coordinate system any more, but only induced by the connection or the covariant derivative.
\bibitem{NoteCurvature}
More precisely, the anholonomy is measured by the variation of the angle after a closed loop $\ga$ in the base space while parallel transported along the connection. This variation is computed by integrating the connection (here $\de\Theta$) along $\ga$. When the loop can be used to define a two-dimensional surface $S$ with boundary $\partial S=\ga$, as is the case when the base space is simply connected (i.e. when it has no non-contractible loops), then the result can be expressed as the integral of the curvature $2$-form on $S$.
\bibitem{NoteMonodromy}
More precisely, making all derivative operators commute only guaranties local existence for the desired scalar coordinates. For a global existence, obstructions could come from non-contractible loops, which could generate a discontinuity or a multi-valuation for the desired coordinate. However, here the base space is simply connected, so that the local criterion is enough to provide global existence.
\end{thebibliography}
\end{document}